\documentclass[journal=jctc,manuscript=article]{achemso}

\usepackage{chemformula} 
\usepackage[T1]{fontenc} 

\SectionNumbersOn
\author{Ivan Scivetti}
\email{ivan.scivetti@stfc.ac.uk}
\affiliation{Daresbury Laboratory, Sc. Tech. Keckwick Lane, Daresbury, WA4 4AD, Warrington, UK}
\alsoaffiliation{Department of Chemistry, University of Liverpool, Liverpool L69 3BX, UK}
\author{Kakali Sen}
\affiliation{Daresbury Laboratory, Sc. Tech. Keckwick Lane, Daresbury, WA4 4AD, Warrington, UK}
\author{Alin M. Elena}
\affiliation{Daresbury Laboratory, Sc. Tech. Keckwick Lane, Daresbury, WA4 4AD, Warrington, UK}
\author{Ilian Todorov}
\affiliation{Daresbury Laboratory, Sc. Tech. Keckwick Lane, Daresbury, WA4 4AD, Warrington, UK}

\title[title]{Reactive molecular dynamics at constant pressure via non-reactive force fields: extending the Empirical Valence Bond method to the isothermal-isobaric ensemble}
\abbreviations{EVB,NPT,FF}
\begin{document}
\begin{abstract}
The Empirical Valence Bond (EVB) method offers a suitable framework to obtain reactive potentials through the coupling of non-reactive force fields. However, most of the implemented functional forms for the coupling terms depend on complex spatial coordinates, which precludes the computation of the stress tensor for condensed phase systems and prevents the possibility to carry out EVB molecular dynamics in the isothermal-isobaric (NPT) ensemble. In this work, we make use of coupling terms that depend on the energy gaps, defined as the energy differences between the participating non-reactive force fields, and derive an expression for the EVB stress tensor suitable for computations. Implementation of this new methodology is tested for a model of a single reactive malonaldehyde solvated in non-reactive water. Computed densities and classical probability distributions in the NPT ensemble reveals a negligible role of the reactive potential in the limit of low concentrated solutions, thus corroborating the validity of standard approximations customarily adopted for EVB simulations. 
\end{abstract}


\section{Introduction}
Molecular dynamics (MD) simulations offer a powerful computational tool to derive atomistic insight of complex phenomena from organic chemistry and biochemistry to heterogeneous catalysis \cite{karplus2002,hollingsworth2018,grajciar2018}. The interatomic interactions in classical MD simulations are based on force field (FF) descriptors\footnote{A force field is a mathematical construction to model the interactions between atoms without having to compute the electronic Schr\"odinger equation. This construction reduces dramatically the computational effort to obtain energies and forces (and stress tensors for extended systems), allowing the sampling of the phase space up to nano-seconds, depending on the system and the computational resources available.} which allows for very fast computation of the interactions and access to simulate very large systems. Commonly, these FF descriptors have simple functional forms, with parameters either fitted to experimental data or derived from quantum mechanical calculations \cite{gonzalez2011,ballone2014}. In most of the available FF libraries, functional forms and fitted parameters remain unchanged during the course of the MD simulation. In reactive processes, however, the interactions inevitably change due to the breaking and/or formation of chemical species. Thus, standard FFs are not suitable to simulate chemical reactions and they are referred to as non-reactive.\\ 
An alternative to simulate chemical reactions with MD is given by Reactive FFs (RFFs) \cite{vanduin2001,farah2012,senftle2016,yun2017,islam2016,liang2013}, that are designed to model interatomic interactions of multiple states representing different chemical species. The task of designing RFFs, however, is very challenging\footnote{Indeed, designing RFFs requires a high level of expertise to tackle a multi-dimensional problem \cite{liang2013}, where the modelled interactions are often expressed by complicated functional forms with many strongly coupled parameters that are optimized via the use of sophisticated tools \cite{pahari2012,larsson2013,li2013,jaramillo2014,larentzos2015,rice2015,dittner2015}.} and, despite the enormous progress over the last years \cite{farah2012,shin2012}, a general parameterization is not yet available.\\ 
The Empirical Valence Bond (EVB) method \cite{warshel1980,aqvist1993,hartke2015,carpenter2015, duarte2017,zhang2019} offers, instead, a simple general framework to model reactive processes through the coupling of multiple non-reactive FFs, where each FF corresponds to a different chemical state for the system. In this method, a suitable EVB matrix is built using the computed energies of the involved chemical states as well as appropriate coupling terms. Matrix diagonalization at each time step allows computation of reactive energy landscapes that account for the change in chemistry when sampling conformations between the participating, chemically different, states.\\
In contrast to RFFs, the advantage of the EVB method lies in the large availability of standard non-reactive FFs libraries, which has offered an appealing strategy for computational implementation and development over the past four decades \cite{molaris-xg,marelius1998,yamashita2012,amber}. Moreover, despite the tedious initial task to calibrate the coupling terms against reference data, research has demonstrated that these couplings are invariant to the surrounding electrostatics, making it possible to simulate the same reactive unit in different environments \cite{hong2006}. This convenient feature of the EVB method has widely increased its recognition as a powerful tool within the computational chemistry community \cite{duarte2017}. \\  
For condensed phase systems, reported MD simulations with the EVB method (herein, MD-EVB simulations) are conducted either in the microcanonical (NVE) or the canonical (NVT) ensemble. However, MD-EVB simulations at constant pressure and temperature, i.e. using the isothermal-isobaric (NPT) ensemble have not been addressed in the literature. In fact, the standard protocol for EVB simulations in condensed phase is to first consider only one of the possible chemical states of the system (preferably the state with lowest free energy) together with the surrounding, non-reactive, environment and carry out a standard NPT simulation (without EVB) at the target pressure and temperature \cite{kakali2017}. The converged volume is then fixed and the MD-EVB simulation is performed using the NVT ensemble. This procedure appears to be a sensible strategy for very large, homogeneous soft-matter systems. However, for smaller systems or higher concentration of solutes, the validity of this standard protocol to approximate real experimental conditions at constant pressure and temperature has never been corroborated to date. \\
In this work, we demonstrate that the use of the standard formulation to compute the stress tensor cannot be directly applied to derive the components of the EVB stress tensor. We argue that this limitation explains the absence of MD-EVB simulations in the NPT ensemble. In contrast to using complex spatial variables to fit the coupling terms of the EVB matrix, we propose to make use of energy gaps \cite{mones2009}, defined as the energy difference between the non-reactive FFs. With this choice, we derive an expression for the EVB stress tensor suitable for computational implementation, not only offering a solution to an overlooked limitation of EVB but also extending the applicability of MD-EVB simulations to NPT ensembles for the first time. The computational implementation of this new formalism is tested using a model of a solvated reactive malonaldehyde molecule in water. MD-EVB simulations at 300 K and 1 atm are used to quantify the role of the reactive potential in the computed density and classical probability distributions of the energy gaps obtained from the sampling of the configurational space. Results also allow to evaluate the validity of the standard protocol for MD-EVB simulations, while the derived method offers an opportunity to explore new strategies for future implementation and development of the EVB method.\\ 
The structure of this paper is as follows. The fundamentals of the EVB method developed over the years are presented in a convenient notation in section \ref{sec:evb}. In section \ref{sec:evb-stress}, we discuss the limitation of the standard formulation to calculate the EVB stress tensor, and propose a new alternative method. A brief overview of the computational implementation is given in section \ref{sec:implement}. Section \ref{sec:coupling} discusses general aspects of the coupling terms within the framework of the present paper. Details of the model and MD computational setting are provided in section \ref{sec:model}, which is followed by section \ref{sec:results} with the results and discussion. Concluding remarks are finally addressed in section \ref{sec:conclusions}.
\section{The EVB method}\label{sec:evb}
In this section we present the fundamentals of the EVB formalism in a convenient notation. Let us assume an atomic system composed of $N_{p}$ particles with positions described by the set  of vectors $\{\bf R\}$. The non-reactive force field (FF) for the chemical state $(m)$ is described by the configurational energy $E_{c}^{(m)}(\{\bf R\})$ and the set of forces $\vec{F}_{J}^{(m)}(\{\bf R\})$, where the index $J$ runs over the total number of particles. The configurational energy function $E_{c}^{(m)}(\{ {\bf R} \})$ can be generally written as a sum of different terms as follows \cite{dlpoly-manual}
\begin{eqnarray}\label{eq:ene-decomp}
E_{c}^{(m)}(\{ {\bf R} \})&=& [ E^{(m)}_{shell}+E^{(m)}_{teth}+E^{(m)}_{bond}+E^{(m)}_{ang}+E^{(m)}_{dih}+ \nonumber \\
                                   &+& E^{(m)}_{inv}+E^{(m)}_{3body}+E^{(m)}_{4body}+E^{(m)}_{ters}+ \nonumber \\
                                   &+& E^{(m)}_{metal}+ E^{(m)}_{vdw}+ E^{(m)}_{coul}] (\{ {\bf R} \})
\end{eqnarray}
where $E^{(m)}_{shell}$, $E^{(m)}_{teth}$, $E^{(m)}_{bond}$, $E^{(m)}_{ang}$, $E^{(m)}_{dih}$, $E^{(m)}_{inv}$, $E^{(m)}_{3body}$, $E^{(m)}_{4body}$, $E^{(m)}_{ters}$, $E^{(m)}_{metal}$, $E^{(m)}_{vdw}$ and , $E^{(m)}_{coul}$ are the interactions representing core-shell polarization, tethered particles, chemical bonds, valence angles, dihedrals, inversion angles, three-body, four-body, Tersoff, metallic, van der Waals and coulombic contributions, respectively. Following Eq.~(\ref{eq:ene-decomp}), the forces can be expressed using a similar decomposition. In the current notation, we shall use indexes $m$ and $k$ for the chemical states (and FFs), $I$ and $J$ for atoms and Greek letters for Cartesian coordinates. Indexes in parenthesis are used to emphasize the particular chemical state.\\
The purpose of the EVB method is to couple $N_F$ non-reactive force fields to obtain a reactive potential. These FFs are coupled through the Hamiltonian $\hat{H}_{\text{EVB}}$ with a matrix representation $H_{\text{EVB}} \in \mathcal{R}^{N_F \times N_F}$ that has the following components
\begin{equation}\label{eq:evbmatrix}
H^{mk}_{\text{EVB}}(\{\bf{R}\})=\begin{cases} E_{c}^{(m)}(\{{\bf R}\})               \,\,\,\,\,\,\,\,\,\,\,\,\,\,\,\,\,\,  m=k   \\
                                                                   C_{mk}(\epsilon_{mk})                     \,\,\,\,\,\,\,\,\,\,\,\,\,\,\,\,\,\,\,\,\,   m \ne k 
                                              \end{cases}
\end{equation}
where each diagonal element corresponds to the configurational energy $E_{c}^{(m)}(\{ {\bf R} \})$ of the non-reactive FF that models the interactions as if the system was in the chemical state $(m)$, whereas the off-diagonal terms C$_{mk}$ are the couplings between states $m$ and $k$. For convenience in the notation, we shall omit hereinafter the dependence on the set of coordinates $\{{\bf R}\}$ for the particles. Even though there are different possible choices for the coupling terms, in the above definition we have set $C_{mk}$ to depend on $\epsilon_{mk}=E_{c}^{(m)}-E_{c}^{(k)}=-[E_{c}^{(k)}-E_{c}^{(m)}]=-\epsilon_{km}$, where $\epsilon_{mk}$ is commonly referred to as energy gap and defines a possible reaction coordinate for the process \cite{hartke2015,duarte2017,mones2009,warshel1991}. Since the $H_{\text{EVB}}$ matrix is Hermitian by construction and the $C_{mk}$ terms are real, the condition of $C_{mk}=C_{km}$ must be imposed to the off-diagonal elements. Diagonalization of $H_{\text{EVB}}$ leads to $N_F$ possible eigenvalues $\{\lambda_1,...,\lambda_{N_{F}}\}$ with
\begin{equation}\label{eq:Heig}
 H_{\text{EVB}}\Psi_{\lambda_m}=\lambda_m \Psi_{\lambda_m}, \,\,\,\,\,\,\,\,\, m=1,...,N_F.
\end{equation}
The EVB energy, $E_{\text{EVB}}$, is defined as the lowest eigenvalue
\begin{equation}\label{eq:Eevb}
 E_{\text{EVB}}=min(\lambda_1,...,\lambda_{N_F})
\end{equation}
with the corresponding normalized EVB eigenvector
\begin{equation}\label{eq:Psi-evb-norm}
 \Psi_{\text{EVB}}=\Psi_{min(\lambda_1,...,\lambda_{N_F})}.
\end{equation}
and
\begin{equation}\label{eq:EevbPsi}
 E_{\text{EVB}}=\big\langle \Psi_{\text{EVB}}\big|\hat{H}_{\text{EVB}}\big| \Psi_{\text{EVB}}\big \rangle.
\end{equation}
Since the eigenvector $\Psi_{\text{EVB}}$ is real and normalized we have
\begin{equation}\label{eq:evbPsinorm}
\sum_{k=1}^{N_F} \big|\Psi^{(k)}_{\text{EVB}}\big|^{2}=1
\end{equation}
from which we can interpret $|\Psi^{(k)}_{\text{EVB}}\big|^{2}$ as the fraction of the chemical state $(k)$ being part of the EVB state. The eigenvector $\Psi_{\text{EVB}}$ can also be represented as a column vector $\in \mathcal{R}^{N_F \times 1}$ where  $\Psi^{(k)}_{\text{EVB}}$ is the element of the $k$-row. Thus, Eq.~(\ref{eq:EevbPsi}) is expressed as a matrix multiplication
\begin{equation}\label{eq:EevbPsimat}
E_{\text{EVB}}=\sum_{m,k=1}^{N_F} \tilde{\Psi}^{(m)}_{\text{EVB}} H^{mk}_{\text{EVB}}\Psi^{(k)}_{\text{EVB}}
\end{equation}
where $\tilde{\Psi}_{\text{EVB}}$ is the transpose of ${\Psi}_{\text{EVB}}$. In section S1 of the Supporting Information we demonstrate that the decomposition of $E_{\text{EVB}}$ into different types of interactions (bonds, angles, etc) as for $E_{c}^{(m)}$ in Eq.~(\ref{eq:ene-decomp}) is not well defined.\\
The resulting EVB force over the particle $J$,  $\vec{F}_{J}^{\text{EVB}}$, follows from the Hellman-Feynman theorem \cite{feynman1939}
\begin{eqnarray}\label{eq:Fevb}
&&\vec{F}_{J}^{\text{EVB}}=-\nabla_{\vec{R}_J}E_{\text{EVB}}=-\big\langle \Psi_{\text{EVB}}\big| \nabla_{\vec{R}_J} \hat{H}_{\text{EVB}} \big| \Psi_{\text{EVB}}\big \rangle \nonumber \\
&&= \sum_{\alpha=x,yz} F_{J\alpha}^{\text{EVB}} \,\, \check{\alpha}
\end{eqnarray}
where $\check{\alpha}$ corresponds to each of the orthonormal Cartesian vectors and 
\begin{equation}\label{eq:Fevb2}
F_{J\alpha}^{\text{EVB}}=-\big\langle \Psi_{\text{EVB}}\big| \frac{\partial \hat{H}_{\text{EVB}}}{\partial_{R_{J\alpha}}}\big| \Psi_{\text{EVB}}\big \rangle.
\end{equation}
From Eq.~(\ref{eq:evbmatrix}) the matrix components of the operator 
$\frac{\partial \hat{H}_{\text{EVB}}}{\partial_{R_{J\alpha}}}$ are given as follows
\begin{equation}\label{eq:gradevb}
\frac{\partial H^{mk}_{\text{EVB}}}{\partial R_{J\alpha}}
=\begin{cases}
\frac{\partial E_{c}^{(m)}}{\partial R_{J\alpha}}=-F^{(m)}_{J\alpha} \,\,\,\,\,\,\,\,\,\,\,\,\,\,\,\,\,\,\,\,\,\,\,\,\,\,\,\,\,\,\,\,\,\,\,\,\,\,\,\,\,\,\,\,\,\,\,\,\,\,\,\,\,\,\,\,\,\,\,\,\,\,\,\,\,\,\,\,\,\,\, m=k  \\
\\
\begin{aligned}
\frac{d C_{mk}}{\partial R_{J\alpha}} &=\frac{d C_{mk}(\epsilon_{mk})}{d\epsilon_{mk}}\frac{\partial \epsilon_{mk}}{\partial R_{J\alpha}}\,\,\,\,\,\,\,\,\,\,\,\,\,\,\,\,\,\,\,\,\,\,\,\,\,\,\,\,\,\,\,\,\,\,\,\,\,\,\,\,\,\,\,  m \ne k\\
                                                          &=\frac{d C_{mk}(\epsilon_{mk})}{d\epsilon_{mk}} \left[\frac{\partial E_{c}^{(m)}}{\partial J\alpha}-\frac{\partial E_{c}^{(k)}}{\partial J\alpha}\right]\\
                                                          &=C^{\prime}_{mk}[F^{(k)}_{J\alpha}-F^{(m)}_{J\alpha}] 
\end{aligned}
\end{cases} 
\end{equation}
where $C^{\prime}_{mk}=\frac{d C_{mk}(\epsilon_{mk})}{d\epsilon_{mk}}$ and  $F^{(k,m)}_{J\alpha}$ is the $\alpha$ component of the total configurational force over particle $J$ in the chemical state $(k,m)$. Similarly to Eq.~(\ref{eq:EevbPsimat}), Eq.~(\ref{eq:Fevb2}) can be expressed as a matrix multiplication
\begin{equation}\label{eq:FevbPsimat}
F_{J\alpha}^{\text{EVB}}=-\sum_{m,k=1}^{N_F} \tilde{\Psi}^{(m)}_{\text{EVB}} \left(\frac{\partial H^{mk}_{\text{EVB}}}{\partial R_{J\alpha}}\right) \Psi^{(k)}_{\text{EVB}}.
\end{equation}	

The above equations define the standard EVB force field (EVB-FF). Even though the EVB formalism was first developed to compute molecular systems, EVB is also applicable to extended systems, customarily modelled using the supercell approximation and periodic boundary conditions (PBCs). Nevertheless, MD-EVB simulations have only been conducted for the NVE and NVT ensembles, to the best of our knowledge, as there is no evidence of a previously reported method to compute the EVB stress tensor. In the next section, we discuss the intricacies related to computing the stress tensor using the standard formulation and propose a new method that allows extending the applicability of MD-EVB to NPT ensembles the first time.

\section{The EVB stress tensor}\label{sec:evb-stress}
The key requirement for a NPT simulation with the EVB method is to being able to compute the EVB stress tensor $\sigma^{\text{EVB}}$.  Similarly to the energy and forces, the configurational stress tensor for the force field $m$, ${\bf \sigma}^{c(m)}$, can be decomposed in a general expression equivalent to Eq.~(\ref{eq:ene-decomp}), where each contribution is computed separately using well-known functional forms \cite{smith1987,dlpoly-manual}. For $bonded$ interactions, for example, the $\alpha\beta$ contribution to the stress tensor from particle $J$ due to the bonded interactions with the surrounding particles, $\sigma_{J,\alpha\beta}^{\text{\tiny{bond}(m)}}$, is given by
\begin{equation}\label{eq:stressbond}
\sigma_{J,\alpha\beta}^{\text{\tiny{bond}}(m)}=\sum_{I}{R}_{JI,\alpha}\,\,{f}^{\text{\tiny{bond}}(m)}_{IJ,\beta}
\end{equation}

where ${R}_{JI,\alpha}$ is the $\alpha$ component of the vector separation $\vec{R}_{JI}=\vec{R}_J-\vec{R}_I$ between particles $I$ and $J$, and $\vec{f}^{\text{\tiny{ bond}}(m)}_{IJ}$ the bond force over particle $J$ from its bonded interaction with particle $I$. In Eq.~(\ref{eq:stressbond}) the sum runs over all particles $I$ interacting with particle $J$ via bonds. Analogously, we could in principle propose an expression for the $\alpha\beta$ component of the EVB stress tensor resulting from the EVB bonded forces, ${f}^{\text{EVB}}_{IJ,\beta}$, as follows,
\begin{equation}\label{eq:stressbondEVB}
\sigma_{J,\alpha\beta}^{\text{EVB}}={R}_{JI,\alpha}\,\,{f}^{\text{EVB}}_{IJ,\beta}.
\end{equation}

In the present case of bonded interactions, the evaluation of Eq.~(\ref{eq:stressbondEVB}) requires of each individual EVB-bonded force over particle $J$ from interaction with particles $I$, given by $\vec{f}^{\text{ EVB}}_{IJ}$. Nevertheless, the EVB force given in Eq.~(\ref{eq:Fevb2}) represents the total force, $\vec{F}_{J}^{\text{ EVB}}$, resulting from the interaction of particle $J$ with all the neighboring particles, which generally include other type of interactions apart from bonding interactions. As far as we can discern, each individual contribution to the force $\vec{f}^{\text{ EVB}}_{IJ}$ cannot be computed from the EVB formalism presented in last section and, consequently, the evaluation of the stress tensor via Eq.~(\ref{eq:stressbondEVB}) is not possible. The same reasoning applies to other type of interactions. This limitation precludes the computation of the stress tensor within the EVB formalism via standard formulae and, consequently, MD simulations using the NPT ensemble. Surprisingly, this inherent limitation of the EVB method has not been previously discussed in the literature, to the best of our knowledge.\\
To circumvent this problem, we propose to use the well-known relation between the configurational energy and the configurational stress tensor \cite{essmann1995}
\begin{equation}\label{eq:stress-def1}
\frac{\partial E^{(k)}_{c}}{\partial h_{\alpha\beta}}=-V\sum_{\gamma=x,y,z}\sigma_{\alpha\gamma}^{c(k)}h^{-1}_{\beta\gamma}
\end{equation}
where $h$ is the set of lattice vectors of the supercell with volume $V$=det($h$). Multiplying to the left by $h_{\nu\beta}$ and summing over $\beta$ we obtain the inverse relation to Eq.~(\ref{eq:stress-def1})
\begin{equation}\label{eq:stress-def2}
\sigma_{\alpha\beta}^{c(k)}=-\frac{1}{V}\sum_{\gamma=x,y,z}h_{\beta\gamma}\frac{\partial E^{(k)}_{c}}{\partial h_{\alpha\gamma}}
\end{equation}
which can be used to define the EVB stress tensor
\begin{equation}\label{eq:stress-def3}
\sigma_{\alpha\beta}^{\text{EVB}}=-\frac{1}{V}\sum_{\gamma=x,y,z}h_{\beta\gamma}\frac{\partial E_{\text{EVB}}}{\partial h_{\alpha\gamma}}.
\end{equation}
Similar to the definition of the EVB force, we evaluate $\partial E_{\text{EVB}}/\partial h_{\alpha\gamma}$ using the Eq.~(\ref{eq:EevbPsi}) and the Hellman-Feynman theorem \cite{feynman1939} 
\begin{equation}\label{eq:stress-EVB}
\frac{\partial E_{\text{EVB}}}{\partial h_{\alpha\beta}}=\big\langle \Psi_{\text{EVB}}\big| \frac{\partial \hat{H}_{\text{EVB}}}{\partial h_{\alpha\beta}}\big| \Psi_{\text{EVB}}\big \rangle.
\end{equation}
The matrix components of the operator $\frac{\partial \hat{H}_{\text{EVB}}}{\partial_{h_{\alpha\beta}}}$ follow from the definition of the EVB matrix (\ref{eq:evbmatrix}) and the use of relation (\ref{eq:stress-def1})
\begin{equation}\label{eq:stress-EVB-mat}
\frac{\partial H^{mk}_{\text{EVB}}}{\partial h_{\alpha\beta}}=\begin{cases}
  \frac{\partial E_{c}^{(m)}}{\partial h_{\alpha\beta}}=-V\sum_{\gamma}\sigma_{\alpha\gamma}^{c(m)}h^{-1}_{\beta\gamma} \,\,\,\,\,\,\,\,\,\,\,\,\,\,\,\,\,\,\,\,\,\,\,\,\,\,\,\,\,\,\,\,\,\,\,\,\,\,\,\,\, m=k  \\
\\
\begin{aligned}
\frac{d C_{mk}}{\partial h_{\alpha\beta}}&= \frac{d C_{mk}(\epsilon_{mk})}{d \epsilon_{mk}}\frac{\partial \epsilon_{mk}}{\partial h_{\alpha\beta}}
  \,\,\,\,\,\,\,\,\,\,\,\,\,\,\,\, \,\,\,\,\,\,\,\, \,\,\,\,\,\,\,\, \,\,\,\,\,\,\,\,    m \ne k
\\ 
                                                              &= \frac{d C_{mk}(\epsilon_{mk})}{d\epsilon_{mk}}\left[\frac{\partial E_{c}^{(m)}}{\partial h_{\alpha\beta}}-\frac{\partial E_{c}^{(k)}}{\partial h_{\alpha\beta}} \right]\
\\
                                                              &=-VC^{\prime}_{mk}\sum_{\gamma}[\sigma_{\alpha\gamma}^{c(m)}-\sigma_{\alpha\gamma}^{c(k)}] h^{-1}_{\beta\gamma}.\\
\end{aligned}
\end{cases} \nonumber 
\end{equation}
Finally, the EVB stress tensor of Eq.~(\ref{eq:stress-def3}) can be expressed as a matrix multiplication
\begin{equation}\label{eq:stress-EVB-ab}
\sigma_{\alpha\beta}^{\text{EVB}}=-\frac{1}{V}\sum_{\gamma=x,y,z}h_{\beta\gamma}\sum_{m,k=1}^{N_F} \tilde{\Psi}^{(m)}_{\text{EVB}} \left(\frac{\partial H^{mk}_{\text{EVB}}}{\partial h_{\alpha\beta}}\right) \Psi^{(k)}_{\text{EVB}}.
\end{equation}
These expressions provide an alternative to compute the stress tensor $\sigma^{\text{EVB}}$ from the configurational stress tensors of each non-reactive FF, $\sigma_{\alpha\gamma}^{c(k)}$. It is important to note that this new scheme to compute $\sigma^{\text{EVB}}$ can only be derived if one uses functional forms for $C_{mk}$ that depend on the energy differences $\epsilon_{mk}$, for which one can evaluate $\frac{\partial E_{c}^{(m)}}{\partial h_{\alpha\beta}}-\frac{\partial E_{c}^{(m)}}{\partial h_{\alpha\beta}}$ and use relation (\ref{eq:stress-def1}) with the computed configurational stress tensor for each chemical state. In contrast, if the choice was to use coupling terms that do not depend on $\epsilon_{mk}$ but other degrees of freedom such as spatial coordinates (see refs. \citenum{chang1990,truhlar2000,schlegel2006,sonnenberg2007,sonnenberg2009,steffen2017,steffen2019}), we cannot discern a clear logic to derive an expression for $\sigma^{\text{EVB}}$, which might explain the fact there is no evidence of any previous reported method to compute the stress tensor using EVB. \\
So far we have presented an alternative to compute the stress tensor $\sigma_{\alpha\beta}^{\text{EVB}}$ but have not discussed the total virial $\mathcal{V}_{\text{EVB}}$. Similarly to the stress tensor, the inability to compute individual contributions of the EVB force prevents the evaluation of the virial using the standard formulation \cite{dlpoly-manual}, and the usual decomposition of the virial depending of the type of interaction under consideration. Within the presented formalism, we compute the virial $\mathcal{V}_{\text{EVB}}$ from $\sigma_{\alpha\beta}^{\text{EVB}}$ as follows
\begin{equation}\label{eq:virial-total}
\mathcal{V}_{\text{EVB}}=-\sum_{\alpha=x,y,z} \sigma_{\alpha\alpha}^{\text{EVB}}.
\end{equation}
In contrast to the EVB energy, it is possible to decomposed the virial into different type of interactions, as we discuss in section S2 of the Supporting Information. The total stress tensor, $\sigma^{T}$, is given by the following general expression
\begin{equation}\label{eq:stress-total}
\sigma^{T}=\sigma^{\text{kin}}+\sigma^{\text{EVB}}+\sigma^{\text{RB}}+\sigma^{\text{bc}}
\end{equation}
where $\sigma^{\text{kin}}$, $\sigma^{\text{RB}}$ and $\sigma^{\text{bc}}$ are the contributions to the stress tensor from the kinetic energy, rigid bodies (RB) and bond constraints (bc), respectively. The EVB method only accounts for the configurational interactions, as described. The kinetic stress tensor is computed as usual from the instantaneous velocities of the particles \cite{dlpoly-manual}. For a particle that is part of a rigid body, the only possible interactions are intermolecular non-bonded interactions (such as coulombic and van der Waals interactions) with other neighboring particles that are not part of the same rigid body. Following the computation of the EVB forces via Eq.~(\ref{eq:Fevb2}), the contribution to the stress from the rigid bodies follows from refs. \citenum{smith1987} and \citenum{essmann1995}
\begin{equation}\label{eq:stress-RG}
\sigma_{\alpha\beta}^{\text{RB}}=\sum_{\mathcal{B}=1}^{N_{\text{RB}}}\sum_{I=1}^{\eta_{\mathcal{B}}} {F}_{I_{\mathcal{B}},\alpha}^{\text{EVB}} d_{I_{\mathcal{B}},\beta}	
\end{equation}
where $\vec{F}_{I_{\mathcal{B}}}$ is the total force over particle $I$ of rigid body $\mathcal{B}$ and $\vec{d}_{I_{\mathcal{B}}}$ the vector distance from atom $I_{\mathcal{B}}$ to the center of mass of the rigid body $\mathcal{B}$. In the above expression, index $\mathcal{B}$ runs over all the rigid bodies. Each rigid body is composed of $\eta_{\mathcal{B}}$ particles. Since, by definition, the topology of rigid bodies remain unaltered during the simulation, the use of RBs within in the present framework is meaningful only to model the environment interacting reactive EVB site. A common example is the use of rigidly constrained water molecules to model a solution.\\
Contributions to the stress tensor from bond constraints, $\sigma_{\alpha\beta}^{\text{bc}}$, are obtained using the SHAKE/RATTLE algorithm \cite{ryckaert1977,andersen1983} during the course of the simulation. This algorithm is independent of the EVB formalism, and corrects for the dynamics of the constrained particles. Finally, frozen particles do not contributed to the stress tensor and are not considered in the formalism. It is important to note that the topology defined via the setting of RBs, frozen atoms and bond constraints must be the consistent for all the coupled FFs, as they impose well defined conditions for the dynamics. For example, if a group of atoms form a rigid body, they must remain a rigid body independently of chemical state under consideration.
 
\section{Overview of the computational implementation}\label{sec:implement}
The EVB method described in section \ref{sec:evb} and its extension for the computation of the stress tensor (section \ref{sec:evb-stress}) were implemented within the DL\_POLY\_4 code \cite{todorov2006,bush2006}. In the standard format, DL\_POLY\_4 reads the initial coordinates, velocities and forces from the CONFIG file. Each particle is labelled according to its specification in the FIELD file, which contains the information of the FF type and parameters for the interactions between the particles. Settings for the MD simulation are specified in the CONTROL file. Initially, the code was modified to allow i) reading multiple ($N_F$) CONFIG and FIELD files, ii) allocating arrays of dimension $N_F$ for the relevant quantities, iii) checking consistency of specification between all force fields and initial coordinates (including any possible constraint such as rigid bodies), iv) reading EVB settings such as coupling terms and v) preventing the execution if there are MD or FF options that are not consistent with a EVB simulation. With regards to this last point, not all type of interactions in the energy decomposition of Eq.~(\ref{eq:ene-decomp}) are suitable to describe reactive interactions. For example, three-body, four-body, Tersoff and metallic interactions are, by construction, not designed to account for different chemical states. Thus, such interactions should only be used to model the surrounding atomic environment interacting with the EVB site.\\
Regarding the EVB method in itself, modifications to the code required to allow for the computations of energies, forces, stress tensor and virials for each of the $N_F$ force-fields separately. From the computed configurational energy of each FF and the choice of the functional forms for the coupling terms, the EVB matrix (\ref{eq:evbmatrix}) is built and diagonalized, and the lowest eigenvalue and the corresponding vector are assigned to $E_{EVB}$ and $\Psi_{EVB}$, respectively. Matrix (\ref{eq:gradevb}) is computed for each particle's Cartesian components and the resulting EVB force is obtained via the matrix multiplication of Eq.~(\ref{eq:FevbPsimat}). From the stress tensors computed for each FF,  matrix (\ref{eq:stress-EVB-mat}) is built for all the $\alpha\beta$ terms and the $\alpha\beta$ component of the EVB stress tensor obtained via Eq.~(\ref{eq:stress-EVB-ab}), and the total virial from Eq.~(\ref{eq:virial-total}). Such EVB calculations are conducted for each time step taking advantage of the domain decomposition as implemented in DL\_POLY\_4 \cite{todorov2006,bush2006}.\\ 
In this implementation, all the $N_F$ force fields are computed in a loop architecture, i.e. one after the other, before being coupled via the EVB method. This means that all the available processors are used to compute each force-field, in contrast to the alternative strategy of dividing processors for each force field. For extended systems, this choice is convenient given the relative high computational cost of the long range Coulombic part in comparison with all the other contributions to the configurational energy. This loop structure increases the computational time by a multiplicative factor of approximately $N_F$ with respect to the required time to compute only a single force field. 
\section{Coupling terms}\label{sec:coupling}
The quality of EVB method depends on the choice for the coupling terms $C_{mk}$, particularly to reproduce accurate interactions at the intermediate region between chemical states $m$ and $k$ where the change of chemistry occurs. Several sophisticated EVB coupling recipes have been proposed over the years \cite{chang1990,truhlar2000,schlegel2006,sonnenberg2007,sonnenberg2009,steffen2017,steffen2019}. Despite their proven success, these recipes use complex internal (spatial) coordinates to couple the force fields. Here, however, we aim to use functional forms $C_{mk}$ that depend on the energy gaps $\epsilon_{mk}=E^{(m)}_{c}-E^{(k)}_{c}$, because these variables not only constitute a possible generalized reaction coordinate \cite{duarte2017,warshel1991,hartke2015,mones2009} but also allow to compute the EVB stress tensor as described in Sec. \ref{sec:evb-stress}, which is the main purpose of the present work.\\ 
The dependence of coupling terms $C_{mk}$ on energy gaps has been previously investigated by B. Hartke {\it et al.}. \cite{hartke2015}. In their work, Density Functional Theory (DFT)\cite{hohenberg1964,kohn1965} was first used to compute the minimum energy path (MEP) between reactant and product states for the bond breaking-formation of several molecules in gas phase. Via highly accurate DFT-derived FFs \cite{grimme2014,grimme2017} for the involved chemical states $m$ and $k$, the authors computed the coupling terms $C_{mk}$  for selected configurations along the corresponding MEP from the individual energies $E^{(m)}_{c}$ and $E^{(k)}_{c}$ and the reference DFT energy. By plotting $C_{mk}$ as a function of $\epsilon_{mk}$, the data was fitted to constants and Gaussian-type of functions.\\ 
The implementation of such procedure necessarily requires the use of force-fields i) consistent with the level of theory that is used to compute the explicit electronic problem for the reaction and ii) accurate enough far from the reference geometry for which they were fitted. Ultimately, meeting these requirements is a non-trivial challenge, generally impossible in many cases, particularly for large systems. In addition, previous research claimed that for several reactions the resulting EVB energy leads to large errors along the MEP, especially in the transition region where, artificial minima are created in the worst cases. To overcome these limitations, a combination of Gaussian functions were proposed to model the coupling terms \cite{steffen2017}, thus offering a promising route for future calibration and development of EVB potentials.
\section{Model study and MD settings}\label{sec:model}
\begin{figure}[t]
\includegraphics[scale=1.2]{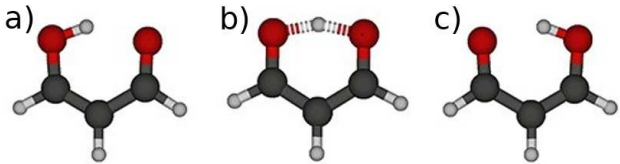}
\centering
\caption{Proton transfer process for malonaldehyde between the two conformations (a and c) via the transition state (b). Oxygen (red), Carbon (dark grey) and hydrogen (light grey).}
\label{fig:model}
\end{figure}
\begin{figure}[t]
\includegraphics[scale=1.2]{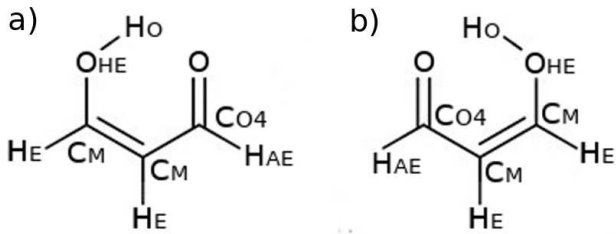}
\centering
\caption{Schematic representation for the two conformations of MA, whose force fields are fitted to model integrations when the proton H$_\text{O}$ is bonded to a) the oxygen at the left b) the oxygen at the right. Note the change of labelling for O, C and H atoms as well as the change of the double bonds. Atom labelling is consistent to the OPLS-2005 FF library.}
\label{fig:label}
\label{fig:model2}
\end{figure}
To test the implementation of EVB method and its extension to the NPT ensemble, we considered a single malonaldehyde molecule in water solution as a our model system. Malonaldehyde (MA) is an archetypal example of intramolecular proton transfer between two oxygen atoms. Each conformation corresponds to a different chemical configurations for the same molecule, as shown in Fig. \ref{fig:model} a and c. In a classical description, the system swaps between both configurations only when vibrations promote the proton to overcome the energy barrier via the transition state (TS), as depicted in Fig. \ref{fig:model}b.\\
A reactive FF for MA would aim to model the interatomic interactions for the whole domain with the forming and breaking of the O-H bonds. An example of such a FF was proposed by Y. Yang {\it et al.}. \cite{yang2010} based on an extension of the molecular mechanics with proton transfer method \cite{lammers2008} to non-linear hydrogen bonds. More recently, reactive force fields for MA have been derived using machine learning \cite{brockherde2017} and neural networks \cite{unke2018}. Here, nevertheless, we use different non-reactive FFs to describe interactions in the vicinity of each conformation, as schematically shown in Fig \ref{fig:label}. The two FFs were generated with the DL\_FIELD program \cite{yong2016} in a format suitable to DL\_POLY\_4 using the OPLS-2005 FF library \cite{banks2005,jorgensen1996}, which is not only specially designed for liquid simulations but also constitutes an example of non-reactive FF available in the literature. Atoms are labelled differently depending on the FF. For example, for the conformation of Fig \ref{fig:label}a (FF$_1$ from now on), the proton H$_\text{O}$ is chemically bonded to the O$_\text{HE}$ site and only interacts with oxygen O via van der Waals and coulombic interactions. For this FF$_1$ topology, the conformational energy would be rather large for geometries where the proton H$_\text{O}$ is at the vicinity of the O site and the realistic chemistry would be better represented by interactions according to the topology of the FF of Fig \ref{fig:label}b (FF$_2$ from now on). For this reason, if one only used FF$_1$ to described the interactions, atom H$_\text{O}$ would unlikely explore the vicinity of the O site during the course of a MD simulation. The same reasoning can be applied to the complementary non-reactive FF$_2$. \\
Each water molecule of the solvent was simulated with the TIP4P scheme \cite{jorgensen1983}, which uses a four-site water model with an off-center point charge for oxygen. To maximize the effect of the EVB reactive potential on the solution, the number of the non-reactive water molecules has to be minimized. In MD simulations with FFs, this choice is restricted by the van der Waals cutoff radius, which is routinely set to 12 \AA. Thus, for a cubic supercell with periodic boundary conditions, the minimum size of the box length should be 24 \AA. To this purpose, models were built to contains 599 rigid water molecules arranged around the MA molecule within a cubic box of 27 \AA, while using an initial separation criteria of 1.9 \AA{} between the molecules. This amount of water prevented box length values below the limit of 24 \AA{} in all the simulations. Such a model already represents an aqueous solution with a rather low concentration of $9.19\times 10^{-2}$ molality [mol(MA)/kg(H$_2$O)].\\
Using the initial arrangement of atoms, the system was initially computed in the NVT ensemble at 300K using only FF$_1$ for MA and a Nose-Hoover \cite{nose1984,hoover1985} thermostat with a relaxation time of 0.5 ps. Equilibration was conducted for 5 ps, scaling the system temperature every 5 fs and resampling instantaneous system momenta distribution every 9 fs. Production MD followed for 30 ps. The last snapshot with positions, velocities and forces served as the starting point of a NPT simulation at 1 atm, this time using a Nose-Hoover thermostat and barostat with relaxation times of 3.0 and 1.0 ps for the thermostat and barostat, respectively. Equilibration was conducted for 2 ps while allowing for a variation of 10\% in the system density. This was followed by a 200 ps of MD production run. The average supercell dimension was used for all the EVB-NVT simulations and as starting point for the EVB-NPT runs. All the MD simulations used a timestep of 1 fs, while the electrostatic interactions are computed through the smooth particle mesh Ewald method \cite{essmann1995,bush2006}. Details of the EVB simulations are given in the next section.
\section{Results and discussion}\label{sec:results}
\begin{figure}[t]
\includegraphics[scale=1.2]{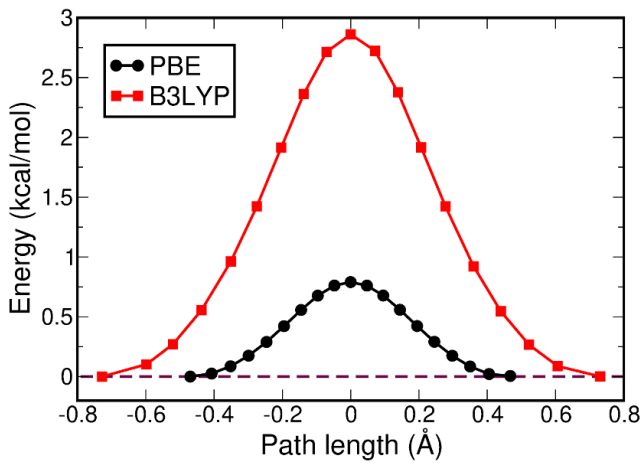}
\centering
\caption{DFT energy profile along the computed MEPs for a single MA molecule in vacuum using the B3LYP (red) and the PBE (black) approximations for the electronic exchange and correlation.}
\label{fig:neb}
\end{figure}
\begin{figure}[t]
\includegraphics[scale=1.2]{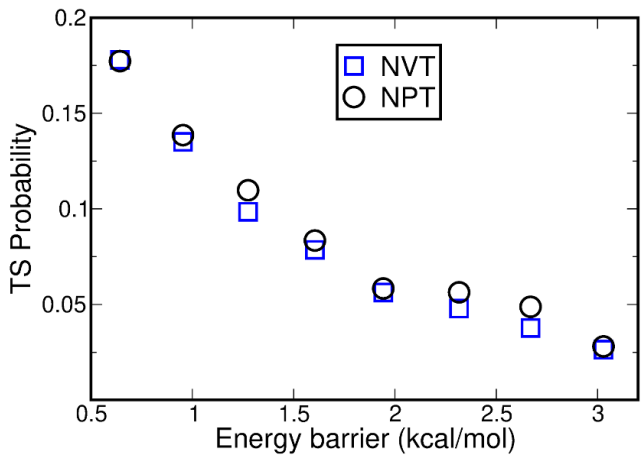}
\centering
\caption{Probability to sample the TS for solvated MA in the NVT and NPT ensembles. Results are plotted as a function of the classical energy barrier for proton transfer in MA.}
\label{fig:TSP}
\end{figure}

Proton transfer is a quantum mechanical process \cite{kiefer2004,marx1999,tuckerman2002}. Accounting for the full quantum problem of the nuclei, however, is computationally prohibited for sufficiently large systems, and several approximations have been developed over the years (See ref. \citenum{yamada2014} and references therein). In particular, the EVB method has been extended to its Multistate version (MS-EVB) to successfully capture the essential physics and chemistry in different protonated systems, both in the classical and quantum regime \cite{schmitt1998,schmitt1999,day2002,wu2008,swanson2007,wick2009,park2012,glowacki2015,biswas2016}. Nevertheless, MS-EVB inherits the limitation of EVB with respect to the stress tensor and its application has only been been restricted to NVE and NVT ensembles.\\  
With regards to the model of a single MA in water, A. Yamada {\it et al.}. has previously used the quantum-classical molecular dynamics method \cite{yamada2008} to compute the quantum reaction dynamics\cite{yamada2014}. One the other hand, Y. Yang {\it et al.}. \cite{yang2010} assumed the whole solution as classical, and used MD to compute proton transfer rates in an effective potential for MA that included zero point energy effects.\\
The purpose of this work, however, is to compute the solvated MA using the EVB method in the NPT ensemble, and compare the results with the standard protocol. Thus, nuclear quantum effects are neglected in the following simulations as well as zero point energy corrections, in line with a previous density-functional tight-binding QM/MM study \cite{walewski2004}. Even though this represents an over simplification of the problem, the assumption of classical mechanics for the whole system has demonstrated to provide a reasonable framework to compute the lower limits for proton transfer \cite{yang2010}.\\
We started our study by considering the explicit quantum electronic problem of a single MA in vacuum at zero temperature, and used the computed quantities as a reference to calibrate the EVB potential. By means of the Nudge Elastic Band (NEB) method \cite{NEB1,NEB2} combined with DFT calculations, we computed the minimum energy path (MEP) to transfer the proton between the two conformations. Details for these calculations are provided in section Settings for DFT simulations. Following the geometry relaxation of each MA conformation, the converged structures were used as fixed end-points of the MEP, which was built by using 17 intermediate images. Fig. \ref{fig:neb} shows the computed energy profiles along the converged MEP for the B3LYP \cite{B3LYP} and the PBE \cite{PBE} electronic exchange and correlation (XC) functionals. Dispersive van der Waals interaction are included via the Grimme's DFT-D3 formalism \cite{grimme2010}. We compute energy barriers of 2.86 and 0.79 kcal/mol for B3LYP and PBE, respectively, which is in agreement with previous work \cite{sadhukhan1999} and corroborates the crucial dependence of the XC functional on the energy barrier and length of the MEP. These values underestimate previous coupled-cluster calculations, which predicted energy barriers between 4.0-4.3 kcal/mol \cite{schroder2011,manthe2011,bowman2008,yang2010}. Despite this underestimation, we shall proceed with the computed DFT values, as achieving chemical accuracy is not the main purpose of the present work. Even though this choice represents a departure from a more realistic chemistry, lower energy barriers are more convenient to MD, as less computational time is needed to switch between conformations, thus allowing a better sampling of the configurational space.\\ 
\begin{figure}[t]
\includegraphics[scale=1.2]{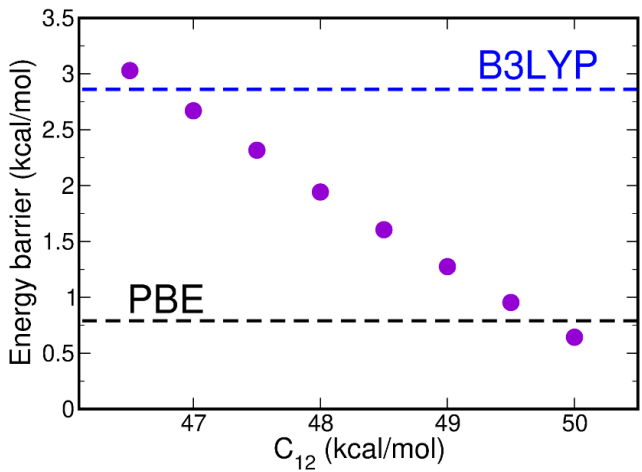}
\centering
\caption{EVB barrier for proton transfer for a MA molecule in vacuum as a function of the coupling term $C_{12}$. Computed energy barriers obtained from the DFT calculations of Fig. \ref{fig:neb} are shown as a reference.}
\label{fig:evb-barrier}
\end{figure}
On the other hand, energy barriers computed with the EVB method depend on the term $C_{12}$. Here, we have assumed $C_{12}$ to be a constant. Note that the choice of a constant for the coupling term, even trivial, complies with the functional form requirement for the coupling terms to compute the EVB stress tensor. Figure \ref{fig:evb-barrier} shows that the computed EVB barrier for MA decreases as the value of $C_{12}$ increases. For the adopted OPLS-2005-FFs and to the purpose of comparison with the DFT energy barriers of Fig. \ref{fig:neb}, we have only considered values of $C_{12}$ in the range between 46.5 and 50.0 kcal/mol. These results constitute an example of how coupling terms can be used to calibrate EVB potentials against a reference value, in this case obtained from DFT. The advantage of the EVB formalism lies in the assumption that the coupling terms calibrated for reactive molecules in gas phase do not change significantly when transferring the reactive system from one phase to the another \cite{duarte2017}. This approximation has been rigorously validated via Constrained-DFT calculation \cite{hong2006}. Moreover, the use of constants for the coupling terms is the most common choice in the execution of EVB simulations for solvated reactive sites, as a constant can be finely adjusted until a calculated property (usually free energy) agrees with the experimental value \cite{duarte2017,kakali2017}. Here, we are not interested in comparing with experiments but evaluating how the results are affected by the use of different ensembles.\\  
Clearly, the reactive EVB potential is different from any of the individual non-reactive FFs, particularly in the TS region. Thus, it is natural to argue as to which extent this reactive EVB potential affects the stress tensor and the converged volume (and density), and how results compare with the density resulting from a NPT simulation using one of the involved FF, as in the standard protocol. If the energy barrier is sufficiently large, the system will only sample the vicinity of one of the possible configurations. Even though the system might occasionally swap conformation, the TS region will be hardly sampled. Thus, for cases where the conformations are chemically equivalent and the barrier is large enough, the standard protocol appears to be a sensible approximation. In contrast, if the barrier is sufficiently low, the TS region will be better sampled during the course of a MD simulation and the average potential will depart from any of the individual FFs. \\
\begin{figure}[t]
\includegraphics[scale=1.2]{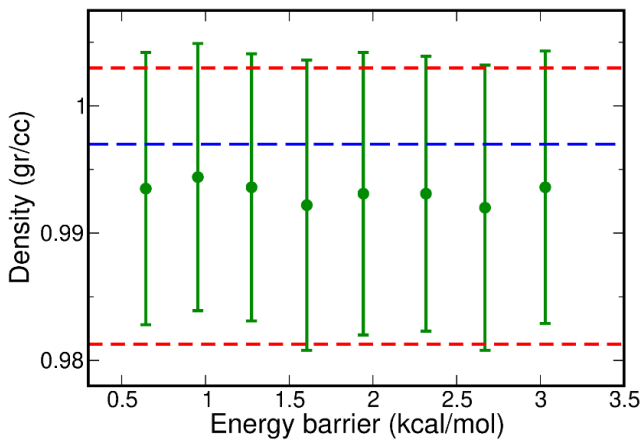}
\centering
\caption{Computed density (green filled circles) of the model solution composed of one MA and 599 rigid water molecules. Different choices of the coupling $C_{12}$ lead to different energy barriers for proton transfer of a single MA in vacuum (see Fig. \ref{fig:evb-barrier}). Each energy barrier (values in the x-axis) can be considered as a different reactive model for MA. The region between horizontal red dashed lines corresponds to the range of possible densities from a standard NPT simulation using only one of the non-reactive FFs. The horizontal blue line refers to the experimental density of pure water. Reference pressure and temperature are 1 atm and 300K, respectively.}
\label{fig:density}
\end{figure}
\begin{figure}[t]
\includegraphics[scale=1.2]{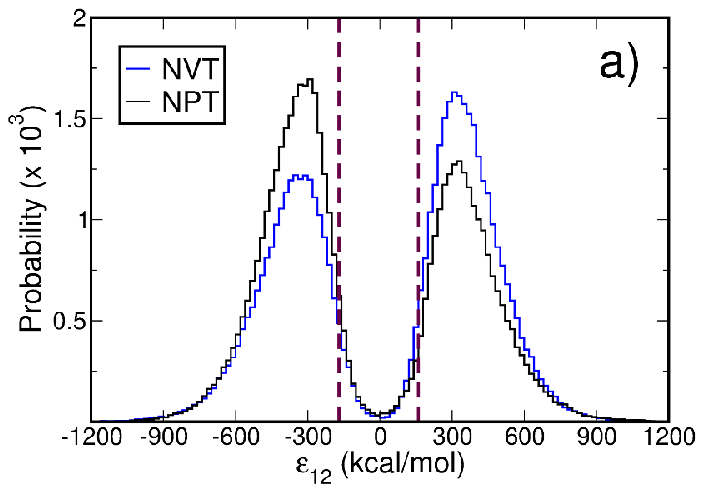}
\includegraphics[scale=1.2]{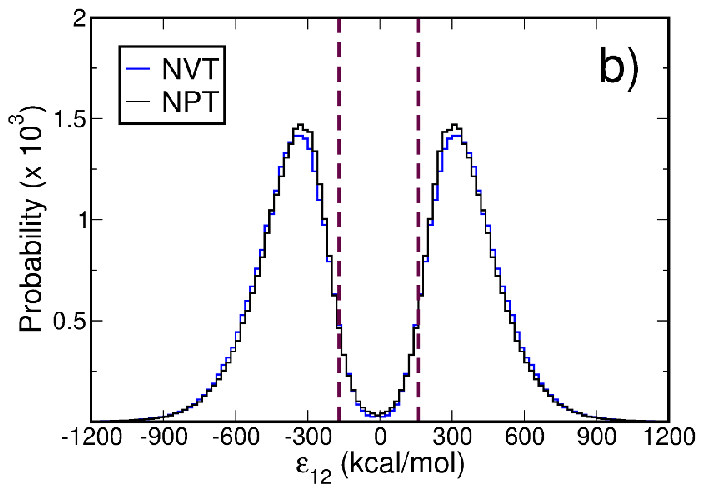}
\centering
\caption{Normalized distribution for the $\epsilon_{12}$ energy gap following NVT and NPT simulations for the model of MA with an energy barrier of 1.94 kcal/mol a) after MD runs b) upon symmetrization. Note values in the y-axis are scaled for the sake of visualization. The region within the brown vertical dashed lines is arbitrarily assigned to the transition state region (see text).}
\label{fig:ene-gap}
\end{figure}
Following a NPT simulation of the solvated MA using only one of the two FFs at 1 atm and 300 K, the computed density of the system is predicted to be in the region between the horizontal red dashed lines, as shown in Fig. \ref{fig:density}. From the set of $C_{12}$  values considered in Fig. \ref{fig:evb-barrier} the classical barrier for proton transfer can be artificially changed. Thus, the EVB potentials generated using these different $C_{12}$  values can be considered as different reactive models for MA. For each of these reactive models, we run a full EVB-NPT simulation and compute the density and its uncertainty, indicated by the green filled circles with error bars. Results demonstrate that the converged density for the solvated MA is statistically independent of the energy barrier for intramolecular proton transfer. In addition, the fact that computed EVB-NPT values statistically fall within the boundary of the red dashed lines supports the validity of the standard protocol in determining the size of the system, as least for the present test case. \\
To further investigate on the role of the ensemble, we use the converged volume from the first NPT simulation (with only one FF) and run EVB-NVT simulations for each reactive field. To compare EVB simulation for both ensembles, here we propose to use the energy gap $\epsilon_{12}$, obtained from the energy difference between the FF$_1$ and FF$_2$ at each time step. The range of computed values for $\epsilon_{12}$ are grouped using a total of 150 bins, each bin with an energy window of 20 kcal/mol. To remove the dependence on the simulation time (i.e. number of configurations sampled), histograms are normalized such as the total area is equal to one. Such normalized distributions can be interpreted as the classical probability of finding the system at a given value of $\epsilon_{12}$. Figure \ref{fig:ene-gap}a shows the computed distribution following EVB-NVT and EVB-NPT simulations for the solvated model with an energy barrier for MA of 1.94 kcal/mol. Distributions exhibit two broad peaks centered at approximately -350 and 350 kcal/mol, which indicates the system mainly samples configurations in the vicinity of the conformations of Figure \ref{fig:model2} and resembles the well-known probability distribution of a proton in a double well. The observed asymmetry in the distributions is attributed to the finite time of MDs run and the lack of control for both configurations to be equally sampled. In fact, the reactive potential for MA in gas phase at zero temperature is symmetric along the MEP (Fig. \ref{fig:neb}), and the same is expected for MA in solution at 300 K despite the electrostatic field created by the surrounding water. Symmetric distributions, however, can only be achieved by increasing the sampling of the conformational phase space and, hence, the computational time for the MD runs, which is beyond the purpose of this work. Alternatively, we use the computed data and make these distributions to be symmetric around $\epsilon_{12}=0$, as shown in Fig \ref{fig:ene-gap}b. To the best of our knowledge, such probability distributions have not been reported before within the framework of EVB. Moreover, these computed distributions support previous research that suggest the convenience of $\epsilon$ as an alternative reaction coordinate \cite{hartke2015,duarte2017,mones2009,warshel1991}. In fact, in contrast to the MEP, $\epsilon$ represents a coordinate that implicitly accounts not only for the reactive site but also includes the effect of the surrounding solvent.\\
The symmetric distributions obtained for the NVT and NPT ensembles indicate a good level of agreement. Nevertheless, it is convenient to quantify this agreement for a better comparison. To this purpose, we first estimate the width for both peaks, which is of the order of 320 kcal/mol. This range for $\epsilon_{12}$ was used to define the TS domain, indicated by the region between the dashed lines of Fig. \ref{fig:ene-gap}, located at $\epsilon^{TS}_{12}= \pm$160 kcal/mol. This TS region is also adopted to be independent of the reactive model for MA. We define the probability for the the system to be in the TS region as follows
\begin{equation}\label{eq:TSP}
\text{TS Probability}(\epsilon^{TS}_{12})=\int_{-\epsilon^{TS}_{12}}^{\epsilon^{TS}_{12}} \mathcal{P}(\epsilon_{12}) \,\, d\epsilon_{12}    
\end{equation}
where $\mathcal{P}(\epsilon_{12})$ is the normalized distribution obtained from the histograms, as shown in Fig. \ref{fig:ene-gap}. Clearly, the computed TS probability depends on the choice for the extension of the TS region, which is completely arbitrary. However, Eq.~(\ref{eq:TSP}) provides a method to quantify the relevance of the TS region and compare the probability distributions for different ensembles. Figure \ref{fig:TSP} shows the computed values for the TS Probabilities for the different reactive models of MA. As expected, the classical probability for the system to sample the TS region increases as the barrier reduces. Additionally, the computed probability distributions do not depend on the assumed ensemble for the present model.\\     
Based on these results, we conclude that the standard protocol to compute single reactive sites in water solution with EVB using the NVT ensemble is already a remarkably good approximation to the self-consistent EVB-NPT simulation, at least for concentrations of reactive sites lower than $9.19\times 10^{-2}$ molality. In the present case study of solvated MA, the flexible-reactive MA is composed of nine atoms and contributes with 27 degrees of freedom (d.o.f), whereas the 599 rigid water molecules can only undergo translation and rotation, contributing another 3594 d.o.f. Thus, the rather small ratio 27:3594 and the rather low compressibility of water might explain why the reactive degrees of freedom play a negligible role in the dynamics of the whole system.\\ 
It would be interesting to investigate if the dynamics will differ on increasing the concentration of the reactive MA. As discussed in section \ref{sec:model}, due to restriction on the size of the simulation cell, higher concentration of MA could only be achieved here by replacing water molecules with MA molecules. However, the EVB method is designed for only one reactive component. Thus, increasing the concentration of reactive components would require an extension of the EVB method to account for multiple reactive sites in a single simulation. This new capability would be ideal not only to the prospect of computing higher concentrations in solutions, but also to perform EVB simulations of molecular crystals composed by reactive units, as for the family of Ketohydrazone-Azoenol systems \cite{gilli2002,durlak2018}. Such a development is part of current research in our group.
\section{Concluding remarks}\label{sec:conclusions}
In this work we propose a new formalism to derive the stress tensor within the EVB method, thus allowing EVB simulations of condensed phase systems at constant pressure and temperature for the first time. This formalism is based on the use of energy gaps as reactive coordinates to parameterize the coupling terms of the EVB matrix. As a test case, we considered the intramolecular proton transfer in MA molecule solvated in water by means of molecular dynamics, while neglecting the role of quantum nuclei and zero point energy corrections.\\
In comparison to the standard protocol of converging the system volume using only one of the non-reactive FFs, results from EVB-NPT simulations for different reactive models of MA (i.e. different energy barriers) demonstrate a negligible effect of the EVB potential in the computed density of the solution.\\ 
In addition, we performed EVB-NVT simulations using the converged volume from the standard protocol. To compare the sampling of the configurational space with respect to EVB-NPT simulations, we use the energy gap as a variable to compute probability distributions of the reactive system. Detailed analysis of these distributions also demonstrate a negligible difference between both ensembles. We attribute these findings to the relative low concentration for the model of MA in water, where the non-reactive dynamics of the rigid water molecules dominates over the reactive dynamics of the single MA. Therefore, future EVB simulations of solvated reactive molecules with higher concentrations would be beneficial to quantify the role of the reactive dynamics on the whole system. To this purpose, developments to extend the EVB method are required to accommodate multiple reactive sites at the same time.  

\section*{Settings for DFT simulations}\label{sec:DFTset}
Geometry optimization and NEB calculations for the two conformations of MA were conducted using the library DL\_FIND \cite{dlfind} of the ChemShell program \cite{chemshell} via its interface \cite{metz2014} with the ORCA package \cite{nesse2012} for the DFT computation of energies and forces. Both PBE and B3LYP functionals were used together with DFT-D3 dispersion correction \cite{grimme2010}. The def2-TZVP basis set \cite{weigend2005} was used for all the atoms.

\section*{Author contributions}\label{sec:contributions}
I.S. and A. M. E. designed the implementation of the EVB in DL\_POLY\_4; I.T has lead DL\_POLY\_4 and participated in initial stages of the EVB project; A. M. E. led the refactoring of DL\_POLY\_4; I.S. developed the mathematical formalism for the EVB stress tensor, implemented the required changes within DL\_POLY\_4, built the models and performed the MD simulations; A. M. E. provided technical support for the EVB implementation; K. S. prepared the input files for the ChemShell simulations; I.S. wrote the manuscript; A. M. E., K. S. and I. T. revised the manuscript.

\begin{acknowledgement}
This work made use of computational support by CoSeC, the Computational Science Centre for Research Communities, through CCP5: The Computer  Simulation of Condensed Phases, EPSRC grant no  EP/M022617/1. I. S. acknowledges i) fruitful discussions with Vlad Sokhan, Silvia Chiacchiera, Fausto Martelli, Alfredo Caro and Gilberto Teobaldi, ii) Chin Yong for his help with the DL\_FIELD code, iii) John Purton for his initial contributions to the EVB funding project iv) Jim Madge for his heroic contribution to the refactoring of DL\_POLY\_4.  A. M. E. acknowledges support of EPSRC via grant EP/P022308/1.
\end{acknowledgement}

\section*{Supporting Information}\label{sec:si}
\subsection*{S1- On the decomposition of the EVB energy}\label{app:evb-decomp}
Similarly to Eq.~(1) of the main manuscript, it is interesting to investigate if $E_{\text{EVB}}$ can be decomposed in different energy contribution. To this purpose we express the diagonal terms of $H^{mk}_{\text{EVB}}$ as a sum of the individual contributions 
\begin{equation}\label{eq:evbmatrix-dec}
H^{mk}_{\text{EVB}}=\begin{cases} \sum_{\text{\tiny{type}}} U_{\text{\tiny{type}}}^{(m)}              \,\,\,\,\,\,\,\,\,\,\,\,\,\,\,\,\,\,  m=k   \\
                                                                     C_{mk}(\epsilon_{mk})                     \,\,\,\,\,\,\,\,\,\,\,\,\,\,\,\,\,\,\,\,\,   m \ne k 
                                              \end{cases}
\tag{S1}
\end{equation}
where the index $\text{type}$ runs over all type of possible interactions (bonds, angles, coulombic, etc). In contrast, $C_{mk}(\epsilon_{mk})$ cannot be decomposed in terms of the individual contributions $U_{\text{\tiny{type}}}^{(m)}$. Consequently, matrix  $H_{\text{EVB}}$ cannot be decomposed for each type of interaction. One might consider the particular case of constant coupling terms $C_{mk}(\epsilon_{mk})=\mathcal{C}_{mk}$, $\forall m,k=1,\cdots, N_F$, with $m\ne k$ to check if a separation into individual terms is possible. For the sake of simplicity, let us consider the case of two FFs with $\mathcal{C}_{12}=\mathcal{C}_{21}$. Without loss of generality, we can write $\mathcal{C}_{12}$ as a sum of a set of constants 
\begin{equation}\label{eq:const-dec}
\mathcal{C}_{12}=\sum_{\text{\tiny{type}}} \mathcal{C}^{\text{\tiny{type}}}_{12}    
\tag{S2}
\end{equation}
and 
\begin{equation}\label{eq:evbmatrix-dec-2x2}
H_{\text{EVB}}=\sum_{\text{\tiny{type}}} H_{EVB}^{\text{\tiny{type}}},\,\,\,\, \text{with} \,\,\,\, H_{EVB}^{\text{\tiny{type}}}=\begin{pmatrix} U_{\text{\tiny{type}}}^{(1)}               & \mathcal{C}^{\text{\tiny{type}}}_{12} \\
                                                                                                                                                                             \mathcal{C}^{\text{\tiny{type}}}_{12} &  U_{\text{\tiny{type}}}^{(2)}\end{pmatrix} 
\tag{S3}
\end{equation}
Using the computed EVB eigenvector, $\Psi_{\text{EVB}}$, from diagonalization of the $H_{\text{EVB}}$ matrix we have
\begin{align}\label{eq:evb_ene-dec-2x2}
E_{\text{EVB}}&=&\sum_{\text{\tiny{type}}} E_{EVB}^{\text{\tiny{type}}}, \,\,\,\,\text{where} \nonumber\\ 
E_{EVB}^{\text{\tiny{type}}}&=&\big\langle \Psi_{\text{EVB}}\big|\hat{H}_{EVB}^{\text{\tiny{type}}}\big| \Psi_{\text{EVB}}\big \rangle 
\tag{S4}
\end{align}
which, in principle, offers a possible way to decompose the EVB energy in terms of individual types of interactions. However, such a decomposition is not unequivocally defined, as there are infinite ways of writing the sum for $\mathcal{C}_{12}$ in Eq.~(\ref{eq:const-dec}). This demonstrates that an EVB energy decomposition in individual terms as in Eq.~(1) is not well defined.  In fact, only $E_{\text{EVB}}$ is well defined.

\subsection*{S2- On the EVB stress tensor and virial decomposition}\label{app:stress-decomp}
The purpose of this section is to demonstrate that the EVB stress tensor $\sigma_{\alpha\beta}^{\text{EVB}}$ and virial $\mathcal{V}_{\text{EVB}}$ can be decomposed in different components according to the type of the interaction. We note that $\sigma_{\alpha\gamma}^{c(m)}$ in Eq.~(19) can be decomposed in different contributions, namely $\sigma_{\alpha\gamma}^{c(m)}=\sum_{\text{\tiny{type}}} \sigma_{\alpha\gamma}^{\text{\tiny{type}}(m)}$. Thus, 
\begin{align}\label{eq:stress-EVB-mat-dec}
&&\frac{\partial H^{mk}_{\text{EVB}}}{\partial h_{\alpha\beta}}=\sum_{\text{\tiny{type}}} \frac{\partial H^{mk}_{\text{\tiny{type}}}}{\partial h_{\alpha\beta}} \,\,\,\,\,\,\,\,\ \text{where}\tag{S5}\\
&&\frac{\partial H^{mk}_{\text{\tiny{type}}}}{\partial h_{\alpha\beta}}=\begin{cases}
-V\sum_{\gamma}\sigma_{\alpha\gamma}^{\text{\tiny{type}}(m)}h^{-1}_{\beta\gamma} \,\,\,\,\,\,\,\,\,\,\,\,\,\,\,\,\,\,\,\,\,\,\,\,\,\,\,\,\,\,\,\,  m=k  \\
\\
-VC^{\prime}_{mk}\sum_{\gamma}[\sigma_{\alpha\gamma}^{\text{\tiny{type}}(m)}-\sigma_{\alpha\gamma}^{\text{\tiny{type}}(k)}] h^{-1}_{\beta\gamma}  \,\,\,\,\,\,\,\,   m \ne k \\
\end{cases} \nonumber 
\end{align}
from which, similarly to Eq.~(20), we have  
\begin{align}\label{eq:stress-type-ab2}
\sigma_{\alpha\beta}^{\text{\tiny{type}}}=-\frac{1}{V}\sum_{\gamma=x,y,z}h_{\beta\gamma}\sum_{m,k=1}^{N_F} \tilde{\Psi}^{(m)}_{\text{EVB}} \left(\frac{\partial H^{mk}_{\text{\tiny{type}}}}{\partial h_{\alpha\beta}}\right) \Psi^{(k)}_{\text{EVB}}\tag{S6}
\end{align}
and
\begin{align}\label{eq:stress-type-ab1}
\sigma_{\alpha\beta}^{\text{EVB}}=\sum_{\text{\tiny{type}}}\sigma_{\alpha\beta}^{\text{\tiny{type}}}
\tag{S7}
\end{align}
with the following decomposition for the virial
\begin{align}\label{eq:virial-decomp}
\mathcal{V}_{\text{EVB}}=\sum_{\text{\tiny{type}}}\mathcal{V}^{\text{\tiny{type}}}_{\text{EVB}}= -\sum_{\text{\tiny{type}}}\sum_{\alpha=x,y,z} \sigma_{\alpha\alpha}^{\text{\tiny{type}}}. \tag{S8}
\end{align}

\bibliography{ms-biblio}

\end{document}